\documentclass{ws-procs9x6}
\usepackage{amsfonts}

\begin{document}

\title{Restricted Complexity, General Complexity\footnote{\uppercase{P}resented at the \uppercase{C}olloquium ``\uppercase{I}ntelligence de la complexit\'e : \'epist\'emologie et pragmatique", \uppercase{C}erisy-\uppercase{L}a-\uppercase{S}alle, \uppercase{F}rance, \uppercase{J}une 26th, 2005". \uppercase{T}ranslated from \uppercase{F}rench by \uppercase{C}arlos \uppercase{G}ershenson.}}
\author{Edgar Morin}
\address{CNRS Emeritus Director\\
Centre d'\'Etudes Transdisciplinaires. Sociologie, Anthropologie, Histoire\\
\'Ecole des Hautes \'Etudes en Sciences Sociales}

\maketitle


Why has the problematic of complexity appeared so late? And why would it be justified?

\section{The three principles of the rejection of complexity by `classical science'}

Classical science rejected complexity in virtue of three fundamental explanatory principles:

\begin{enumerate}
\item The principle of universal determinism, illustrated by Laplace's Daemon, capable, thanks to his intelligence and extremely developed senses, of not only knowing all past events, but also of predicting all events in the future.
\item The principle of reduction, that consists in knowing any composite from only the knowledge of its basic constituting elements.
\item The principle of disjunction, that consists in isolating and separating cognitive difficulties from one another, leading to the separation between disciplines, which have become hermetic from each other.
\end{enumerate}

These principles led to extremely brilliant, important, and positive developments of scientific knowledge up to the point where the limits of intelligibility which they constituted became more important than their elucidations.

In this scientific conception, the notion of ``complexity" is absolutely rejected. On the one hand, it usually means confusion and uncertainty; the expression ``it is complex" in fact expresses the difficulty of giving a definition or explanation. On the other hand, since the truth criterion of classical science is expressed by simple laws and concepts, complexity relates only to appearances that are superficial or illusory. Apparently, phenomena arise in a confused and dubious manner, but the mission of science is to search, behind those appearances, the hidden order that is the authentic reality of the universe.

Certainly, western science is not alone in the search of the ``true" reality behind appearances; for Hinduism, the world of appearances, the m\^ay\^a, is illusory; and for Buddhism the samsara, the world of phenomena, is not the ultimate reality. But the true reality, in the Hindu or Buddhist worlds, is inexpressible and in extreme cases unknowable. Whereas, in classical science, behind appearances, there is the impeccable and implacable order of nature.

Finally, complexity is invisible in the disciplinary division of the real. In fact, the first meaning of the word comes from the Latin \emph{complexus}, which means what is woven together. The peculiarity, not of the discipline in itself, but of the discipline as it is conceived, non-communicating with the other disciplines, closed to itself, naturally disintegrates complexity.

For all these reasons, it is understood why complexity was invisible or illusory, and why the term was rejected deliberately.

\section{Complexity: A first breach: irreversibility}

However, a first breach is made within the scientific universe during the nineteenth century; complexity would appear from it \textit{de facto} before starting to be recognized \textit{de jure}.

Complexity would make its appearance \textit{de facto} with the second law of thermodynamics, which indicates that energy degrades into caloric form: this principle lies within the scope of the irreversibility of time, while until then physical laws were in principle reversible and that even in the conception of life, the fixism of species did not need time.

The important point here is not only the irruption of irreversibility, thus time, but it is also the apparition of a disorder since heat is conceived as the agitation of molecules; the disordered movement of each molecule is unpredictable, except at a statistical scale where distribution laws can be determined effectively.

The law of the irreversible growth of entropy has given place to multiple speculations, and beyond the study of closed systems, a first reflection about the universe, where the second law leads toward dispersion, uniformity, and thus towards death. This conception of the death of the universe, long ago rejected, has appeared recently in cosmology, with the discovery of black energy. This will lead to the dispersion of galaxies and would seem to announce us that the universe tends to a generalized dispersion. As the poet Eliot said: ``the universe will die in a whisper"...

Thus, the arrival of disorder, dispersion, disintegration, constituted a fatal attack to the perfect, ordered, and determinist vision.

And many efforts will be needed---we are not there precisely because it is against the reigning paradigm---to understand that the principle of dispersion, which appears since the birth of the universe with this incredible deflagration improperly named big bang, is combined with a contrary principle of bonding and organization which is manifested in the creation of nuclei, atoms, galaxies, stars, molecules, and life. 

\section{Interaction Order/Disorder/Organization}

How is it that both phenomena are related?

This is what I tried to show in the first volume of \textit{La M\'ethode} (The Method). We will need to associate the antagonist principles of order and disorder, and associate them making another principle emerge that is the one of organization.

Here is in fact a complex vision, which one has refused to consider during a very long time, for one cannot conceive that disorder can be compatible with order, and that organization can be related to disorder at all, being antagonist to it.

At the same time than that of the universe, the implacable order of life is altered. Lamarck introduces the idea of evolution, Darwin introduces variation and competition as motors of evolution. Post-darwinism, if it has, in certain cases, attenuated the radical character of the conflict, has brought this other antinomy of order: chance, I would say even a vice of chance. Within the neodarwinian conception, to avoid calling ``creation" or ``invention" the new forms of living organization such as wings, eyes---one is very afraid of the word ``invention" and of the word ``creation"---one has put chance at the prow. One can understand the rest of the fear of creation because science rejects creationism, 
i.e. the idea that God is a creator of living forms. But the reject of creationism finished in masking the creativity that manifests itself in the history of life and in the history of humanity. And, from the philosophical point of view, it is rather recently that Bergson, and then in another way, Castoriadis, put at the centre of their conception the idea of creation.

In addition, in the beginning of the twentieth century, microphysics introduced a fundamental uncertainty in the universe of particles that ceases to obey the conceptions of space and time characteristic of our universe called macro-physic. How thus these two universes, that are the same, but at a different scale, are compatible? One begins today to conceive that one can pass, from the micro-physical universe to ours, since between them a certain number of quantum elements are connected, in virtue of a process called decoherence. But there remains this formidable logical and conceptual hiatus between the two physics.

Finally, at a very large scale---mega-physical---Einstein's theory discovers that space and time are related to one another, with the result that our lived and perceived reality becomes only meso-physical, situated between micro-physic reality and mega-physical reality

\section{Chaos}

All this made that the dogmas of classical science are reached, but \textit{de facto}: although increasingly mummified, they remain.

Yet a certain number of strange terms would appear. For example, the term ``catastrophe", suggested by Ren\'{e} Thom to try to make intelligible the discontinuous changes of form; then the fractalism of Mandelbrot; then the physical theories of chaos, which distinguishes itself from the rest, since today it is thought that the solar system, which seems to obey an absolutely impeccable and measurable order with the most extreme precision, considering its evolution in millions of years, is a chaotic system comprising a dynamic instability modifying for example Earth's rotation around itself or around the Sun.
A chaotic process may obey to deterministic initial states, but these cannot be know exhaustively, and the interactions developed within this process alter any prevision. Negligible variations have considerable consequences over large time scales. The word chaos, in these physics, has a very limited meaning: that of apparent disorder and unpredictability. Determinism is saved in principle, but it is inoperative since one cannot know exhaustively the initial states. We are in fact, since the original deflagration and forever, plunged in a chaotic universe.

\section{The emergence of the notion of complexity}

However, complexity remained always unknown in physics, in biology, in social sciences. Admittedly, after more than half a century, the word complexity irrupted, but in a domain that also remained impermeable to the human and social sciences, as well as to the natural sciences themselves. It is at the bosom of a sort of nebulous spiral of mathematicians and engineers where it emerged at about the same time, and became connected at once, in the forties and fifties, with Information Theory, Cybernetics, and General Systems Theory. Within this nebula, complexity will appear with Ashby to define the degree of variety in a given system. The word appears, but does not contaminate, since the new thinking remains pretty confined: the contributions of Von Neumann, of Von Foerster will remain completely ignored, and still remain in the disciplinary sciences closed on themselves. One can also say that Chaitin's definition of randomness as algorithmic incompressibility becomes applicable to complexity. Consequently, the terms chance, disorder, complexity tend to overlap one another and sometimes to be confused.

There were breaches, but still not an opening.

This would come from the Santa Fe Institute (1984) where the word will be essential to define dynamical systems with a very large number of interactions and feedbacks, inside of which processes very difficult to predict and control take place, as ``complex systems", where the classical conception was unable to be considered.

Thus, the dogmas or paradigms of classical science began to be disputed.

The notion of emergence appeared. In ``Chance and Necessity", Jacques Monod makes a great state of emergence, i.e. qualities and properties that appear once the organization of a living system is constituted, qualities that evidently do not exist when they are presented in isolation. This notion is taken, here and there, more and more, but as a simple constatation without being really questioned (whereas it is a conceptual bomb).

It is like this that it was arrived to the complexity I call ``restricted": the word complexity is introduced in ``complex systems theory"; in addition, here and there the idea of ``sciences of complexity" was introduced, encompassing the fractalist conception and chaos theory.

Restricted complexity spread rather recently, and after a decade in France, many barriers have been jumped. Why? Because more and more a theoretical vacuum was faced, because the ideas of chaos, fractals, disorder, and uncertainty appeared, and it was necessary at this moment that the word complexity would encompass them all. Only that this complexity is restricted to systems which can be considered complex because empirically they are presented in a multiplicity of interrelated processes, interdependent and retroactively associated. In fact, complexity is never questioned nor thought epistemologically.

Here the epistemological cut between restricted and generalized complexities appears because I think that any system, whatever it might be, is complex by its own nature.

Restricted complexity made it possible important advances in formalization, in the possibilities of modeling, which themselves favor interdisciplinary potentialities. But one still remains within the epistemology of classical science. When one searches for the ``laws of complexity", one still attaches complexity as a kind of wagon behind the truth locomotive, that which produces laws. A hybrid was formed between the principles of traditional science and the advances towards its hereafter. Actually, one avoids the fundamental problem of complexity which is epistemological, cognitive, paradigmatic. To some extent, one recognizes complexity, but by decomplexifying it. In this way, the breach is opened, then one tries to clog it: the paradigm of classical science remains, only fissured.

\section{Generalized complexity}

But then, what is ``generalized" complexity? It requires, I repeat, an epistemological rethinking, that is to say, bearing on the organization of knowledge itself.

And it is a paradigmatic problem in the sense that I have defined ``paradigm"\footnote{Cf \textit{La M\'ethode 4, Les id\'{e}es}, p.211--238,  Le Seuil, 1990}. Since a paradigm of simplification controls classical science, by imposing a principle of reduction and a principle of disjunction to any knowledge, there should be a paradigm of complexity that would impose a principle of distinction and a principle of conjunction.

In opposition to reduction, complexity requires that one tries to comprehend the relations between the whole and the parts. The knowledge of the parts is not enough, the knowledge of the whole as a whole is not enough, if one ignores its parts; one is thus brought to make a come and go in loop to gather the knowledge of the whole and its parts. Thus, the principle of reduction is substituted by a principle that conceives the relation of whole-part mutual implication.

The principle of disjunction, of separation (between objects, between disciplines, between notions, between subject and object of knowledge), should be substituted by a principle that maintains the distinction, but that tries to establish the relation.

The principle of generalized determinism should be substituted by a principle that conceives a relation between order, disorder, and organization. Being of course that order does not mean only laws, but also stabilities, regularities, organizing cycles, and that disorder is not only dispersion, disintegration, it can also be blockage, collisions, irregularities.

Let us now take again the word of Weaver, from a text of 1948, to which we often referred, who said: the XIX$^{th}$ century was the century of disorganized complexity and the XX$^{th}$ century  must be that of organized complexity.

When he said ``disorganized complexity", he thought of the irruption of the second law of thermodynamics and its consequences. Organized complexity means to our eyes that systems are themselves complex because their organization supposes, comprises, or produces complexity.

Consequently, a major problem is the relation, inseparable (shown in \textit{La M\'ethode 1}), between disorganized complexity and organized complexity.

Let us speak now about the three notions that are present, but to my opinion not really thought of, in restricted complexity: the notions of system, emergence, and organization.

\section{System: It should be conceived that ``any system is complex"}

What is a system? It is a relation between parts that can be very different from one another and that constitute a whole at the same time organized, organizing, and organizer.

Concerning this, the old formula is known that the whole is more than the sum of its parts, because the addition of qualities or properties of the parts is not enough to know those of the whole: new qualities or properties appear, due to the organization of these parts in a whole, they are emergent.

But there is also a substractivity which I want to highlight, noticing that the whole is not only more than the sum of its parts, but it is also less that the sum of it parts.

Why?

Because a certain number of qualities and properties present in the parts can be inhibited by the organization of the whole. Thus, even when each of our cells contains the totality of our genetic inheritance, only a small part of it is active, the rest being inhibited. In the human relation individual-society, the possibilities of liberties (delinquent or criminal in the extreme) inherent to each individual, will be inhibited by the organization of the police, the laws, and the social order.

Consequently, as Pascal said, we should conceive the circular relation:  `one cannot know the parts if the whole is not known, but one cannot know the whole if the parts are not known'.

Thus, the notion of organization becomes capital, since it is through organization of the parts in a whole that emergent qualities appear and inhibited qualities disappear\footnote{I develop the idea that organization consists of complexity in \textit{La M\'ethode 1, La nature de la nature}, p.94--151, Le Seuil, 1977.}.

\section{Emergence of the notion of emergence}

What is important in emergence is the fact that it is indeductible from the qualities of the parts, and thus irreducible; it appears only parting from the organization of the whole. This complexity is present in any system, starting with H$_2$O, the water molecule which has a certain number of qualities or properties that the hydrogen or oxygen separated do not have, which have qualities that the water molecule does not have.

There is a recent number of the \textit{Science et Avenir} \footnote{Science and Future, a popular French journal (Translator's Note)} journal devoted to emergence; to relate emergence and organization, one wonders wether it is a hidden force in nature, an intrinsic virtue.

From the discovery of the structure of the genetic inheritance in DNA, where it appeared that life was constituted from physicochemical ingredients present in the material world, therefore from the moment that it is clear that there is not a specifically living matter, a specifically living substance, that there is no \textit{\'elan vital} in Bergson's sense, but only the physicochemical matter that with a certain degree of organizing complexity produces qualities of the living---of which self-reproduction, self-reparation, as well as a certain number of cognitive or informational aptitudes, as from this moment, the vitalism is rejected, the reductionism should be rejected, and it is the notion of emergence that takes a cardinal importance, since a certain type of organizing complexity produces qualities specific of self-organization.

The spirit (\textit{mens, mente}) is an emergence. It is the relation brain-culture that produces as emergent psychic, mental qualities, with all that involves language, consciousness, etc.

Reductionists are unable to conceive the reality of the spirit and want to explain everything starting from the neurons. The spiritualists, incapable of conceiving the emergence of the spirit starting from the relation brain-culture, make from the brain at most a kind of television.

\section{The complexity of organization}

The notion of emergence is a capital notion, but it redirects to the problem of organization, and it is organization which gives consistence to our universe. Why is there organization in the universe? We cannot answer this question, but we can examine the nature of organization.

If we think already that there are problems of irreducibility, of indeductibility, of complex relations between parts and whole, and if we think moreover that a system is a unit composed of different parts, one is obliged to unite the notion of unity and that of plurality or at least diversity. Then we realize that it is necessary to arrive at a logical complexity, because we should link concepts which normally repel each other logically, like unity and diversity. And even chance and necessity, disorder and order, need to be combined to conceive the genesis of physical organizations, as on the plausible assumption where the carbon atom necessary to the creation of life was constituted in a star former to our sun, by the meeting exactly at the same time---absolute coincidence---of three helium nuclei. Thus, in stars where there are billions of interactions and meetings, chance made these nuclei to meet, but when this chance occurs, it is necessary that a carbon atom will be constituted.

You are obliged to connect all these disjoined notions in the understanding that was inculcated to us, unfortunately, since childhood, order, disorder, organization.

We then manage to conceive what I have called the self-eco-organization, i.e. the living organization.

\section{The self-eco-organization}

The word self-organization had emerged and had been used as of the end of the 50's by mathematicians, engineers, cyberneticians, neurologists.

Three important conferences had been held on the topic of ``self-organizing systems", but a paradoxical thing, the word had not bored in biology, and was a marginal biologist, Henri Atlan, who retook this idea, in a great intellectual isolation within his corporation, in the 70's. Finally the word emerged in the 80's-90's in Santa Fe as a new idea, whereas it existed already for nearly half a century. But it is still not imposed in biology.

I call self-eco-organization to the living organization, according to the idea that self-organization depends on its environment to draw energy and information: indeed, as it constitutes an organization that works to maintain itself, it degrades energy by its work, therefore it must draw energy from its environment. Moreover, it must seek its food and defend against threats, thus it must comprise a minimum of cognitive capacities.

One arrives to what I call logically the complex of autonomy-dependence. For a living being to be autonomous, it is necessary that it depends on its environment on matter and energy, and also in knowledge and information. The more autonomy will develop, the more multiple dependencies will develop. The more my computer will allow me to have an autonomous thought, the more it will depend on electricity, networks, sociological and material constraints. One arrives then to a new complexity to conceive living organization: the autonomy cannot be conceived without its ecology. Moreover, it is necessary for us to see a self-generating and self-producing process, that is to say, the idea of a recursive loop which obliges us to break our classical ideas of product $\rightarrow$ producer, and of cause $\rightarrow$ effect.

In a self-generating or self-producing or self-poetic or self-organizing process, the products are necessary for their own production. We are the products of a process of reproduction, but this process can continue only if we, individuals, couple to continue the process. Society is the product of interactions between human individuals, but society is constituted with its emergencies, its culture, its language, which retroacts to the individuals and thus produces them as individuals supplying them with language and culture. We are products and producers. Causes produce effects that are necessary for their own causation.

Already the loop idea had been released by Norbert Wiener in the idea of feedback, negative as well as positive, finally mainly negative; then it was generalized without really reflecting on the epistemological consequences which it comprised. Even in the most banal example which is that of a thermal system supplied with a boiler which provides the heating of a building, we have this idea of inseparability of the cause and effect: thanks to the thermostat, when 20$^\circ$ is reached, the heating stops; when the temperature is too low, the heating is started. It is a circular system, where the effect itself intervenes in the cause which allows the thermal autonomy of the whole compared to a cold environment. That is to say that the feedback is a process which complexifies causality. But the consequences of this had not been drawn to the epistemological level.

Thus feedback is already a complex concept, even in non-living systems. Negative feedback is what makes it possible to cancel the deviations that unceasingly tend to be formed like the fall in temperature compared to the standard. Positive feedback develops when a regulation system is not able anymore to cancel the deviations; those can then be amplified and go towards a runaway, kind of generalized disintegration, which is often the case in our physical world. But we could see, following an idea advanced more than fifty years ago by Magoroh Maruyama, that the positive feedback, i.e. increasing deviation, is an element that allows transformation in human history.  All the great transformation processes started with deviations, such as the monotheist deviation in a polytheist world, the religious deviation of the message of Jesus within the Jewish world, then, deviation in the deviation, its transformation by Paul within the Roman empire; deviation, the message of Mohammed driven out of Mecca, taking refuge in Medina. The birth of capitalism is itself deviating in a feudal world. The birth of modern science is a deviating process from the XVII$^{th}$ century. Socialism is a deviating idea in the XIX$^{th}$ century. In other words, all the processes start by deviations that, when they are not suffocated, exterminated, are then able to make chain transformations.

\section{The relationship between local and global}

In \textit{logical} complexity, you have the relation between the local and the global.

One believed to be able to assume the two truths of the global and of the local with axioms of the style: ``think globally and act locally". In reality, one is, I believe, constrained in our global age to think jointly locally and globally and to try to act at the same time locally and globally. Also, which is also complex, local truths can become global errors. For example, when our immune system rejects with the greatest energy the heart that one grafts to him, like a nasty foreigner, this local truth becomes a global error, because the organism dies. But one can also say that global truths can lead to local errors. The truth of the need to fight against terrorism can lead to interventions, which will favor even more the development of terrorism, just look at Irak.

\section{Heraclitus: ``live of death, die of life"}

In this union of notions \textit{logically} complex, there is a relationship between life and death.

I often quoted the illuminating phrase of Heraclitus, from the VI$^{th}$ century b.C.:  ``live of death, die of life". It became recently intelligible, from the moment when we learned that our organism degrades its energy, not only to reconstitute its molecules, but that our cells themselves are degraded and that we produce new cells. We live from the death of our cells. And this process of permanent regeneration, almost of permanent rejuvenilization, is the process of life. What makes it possible to add to the very right formula of Bichat, saying: ``life is the ensemble of the functions that fight against death", this strange complement that presents us a logical complexity: ``Integrating death to fight better against death". What one again knows about this process is extremely interesting: it has been learned rather recently that cells that die are not only old cells; in fact apparently healthy cells receiving different messages from neighboring cells, ``decide", at a given moment, to commit suicide. They commit suicide and phagocytes devour their remains. Like this, the organism determines which cells must die before they have reached senescence. That is to say that the death of cells and their postmortem liquidation are included in the living organization.

There is a kind of phenomenon of self-destruction, of apoptosis, since this term has been taken from the vegetal world, indicating the split of the stems operated by trees in autumn so that dead leafs fall.

On the one hand, when there is an insufficiency of cellular deaths following different accidents and perturbations, there are a certain number of diseases that are deadly in the long run, like osteoporosis, various types of sclerosis, and certain cancers, where cells refuse to die, becoming immortal, forming tumors and go for a stroll in the form of metastases (It can seem that it is a revolt of cells against their individual death that lead to these forms of death of the organism). On the other hand, the excess of cellular deaths determine AIDS, Parkinson's, and Alzheimer's disease.

You see at which point this relationship between life and death is complex: it is necessary for cells to die, but not too much! One lives between two catastrophes, the excess or insufficiency of mortality. One finds again the fundamentally epistemological problem of generalized complexity.

\section{On non-trivial machines}

Living beings are certainly machines, but unlike artificial machines that are trivial deterministic machines (where one knows the outputs when one knows the inputs), these are non-trivial machines (von Foerster) where one can predict innovative behaviors.

We are machines, this truth was already in \textit{L'homme-machine} of La Mettrie. We are physical machines, thermal machines, we function at the temperature of 37$^\circ$. But we are complex machines.

Von Neumann established the difference between living machines and artificial machines produced by technology: the components of the technical machines, having the good quality of being extremely reliable, go towards their degradation, towards their wear, from the very start of their operation. Whereas the living machine, made up mainly by components far from reliable, degrading proteins---and one understands very well that this lack of reliability of proteins makes it possible to reconstitute them non stop---is able to be regenerated and repaired; it also goes towards death, but after a process of development. The key of this difference lies in the capacity of self-repair and self-regeneration. The word regeneration is capital here.

One can say that the characteristic of innovations that emerge in the evolution of life (which are determined by environmental changes, or by the irruption of multiple hazards), such as the appearance of the skeleton in vertebrates, wings in insects, birds, or bats, all these creations, are characteristic non-trivial machines. That is to say, it gives a new solution to insurmountable challenges without this solution.

All the important figures of human history, on the intellectual, religious, messianic, or politic levels, were non-trivial machines. One can advance that all the History of Humankind, which begins ten thousand years ago---is a non-trivial history, i.e. a history made of unforeseen, of unexpected events, of destructions and creations. The history of life that precedes it is a non-trivial history, and the history of the universe, where the birth of life and then of humankind are included, is a non-trivial history.

We are obliged to detrivialize knowledge and our worldview.

\section{To complexify the notion of chaos}

We have seen how the notion of system brings us to complexities of organization which themselves lead us to logical complexities. Let us look now at the notion of chaos, as it appears within chaos theory, and which comprises disorder and impredictibility. The beat of the wings of a butterfly in Melbourne can cause by a succession of chain processes a hurricane in Jamaica, for example.

Actually, I believe that the word chaos must be considered in its deep sense, its Greek sense. We know that in the Greek worldview, Chaos is at the origin of Cosmos. Chaos is not pure disorder, it carries within itself the indistinctness between the potentialities of order, of disorder, and of organization from which a cosmos will be born, which is an ordered universe. The Greeks saw a bit too much order in the cosmos, which is effectively ordered because the immediate spectacle, the impeccable order of the sky that we see each night with the stars, is always in the same place. And if the planets are mobile they also come to the same place with an impeccable order. However, we know today with the widened conceptions of cosmic time that all this order is at the same time temporary and partial in a universe of movement, collision, transformation.

Chaos and Cosmos are associated---I have employed the word Chaosmos---there is also a circular relation between both terms. It is necessary to take the word chaos in a much deeper and more intense sense than that of physical chaos theory.

\section{The need of contextualization}

Let us take again the ``complexus" term in the sense of ``what is woven together".

It is a very important word, which indicates that the breaking up of knowledge prevents from linking and contextualizing.

The knowledge mode characteristic of disciplinary science isolates objects, one from another, and isolates them compared to their environment. One can even say that the principle of scientific experimentation allows to take a physical body in Nature, to isolate it in an artificial and controlled laboratory environment, and then study this object in function of perturbations and variations that one makes it undergo. This indeed makes it possible to know a certain number of its qualities and properties. But one can also say that this principle of decontextualization was ill-fated, as soon as it was ported to the living. The observation since 1960 by Jane Goodall of a tribe of chimpanzees in their natural environment could show the supremacy of observation (in a natural environment) over experimentation (in a laboratory) for knowledge\footnote{See ``\textit{Le Paradigme Perdu}", pp. 51--54. }. A lot of patience was necessary so that Jane Goodall could perceive that chimpanzees had different personalities, with rather complex relations of friendship, of rivalry; a whole psychology, a sociology of chimpanzees, invisible to the studies in a laboratory or in a cage, appeared in their complexity.

The idea of knowing the living in their environment became capital in animal ethology. Let us repeat it, the autonomy of the living needs to be known in its environment.

From now on, becoming aware of the degradations that our techno-economic development makes to the biosphere, we realize the vital link with this same biosphere that we believe to have reduced to the rank of manipulable object. If we degrade it, we degrade ourselves, and if we destroy it, we destroy ourselves.

The need for contextualization is extremely important. I would even say that it is a principle of knowledge: Anybody who has made a translation in a foreign language will seek an unknown word in the dictionary; but with words being polysemous, it is not immediately known which is the good translation; the sense of the word will be sought in the sense of the sentence in the light of the global sense of the text. Though this play from text to word, and from text to context, and from context to word, a sense will crystalize. In other words, the insertion in the text and in the context is an evident cognitive necessity. Take for example the economy, the most advanced social science from a mathematical point of view, but which is isolated from human, social, historic, and sociologic contexts: its prediction power is extremely weak because the economy does not function in isolation: its forecasts need to be unceasingly revised, which indicates us the disability of a science that is very advanced but too closed.

More generally, mutual contextualization is lacking in the whole of social sciences.

I have often quoted the case of the Aswan dam because it is revealing and significant: it was built in Nasser's Egypt because it would make it possible to regulate the course of a capricious river, the Nile, and to produce electric power for a country which had a great need for it. However, after some time, what happened? This dam retained a part of the silts that fertilized the Nile valley, which obliged the farming population to desert the fields and overpopulate large metropolises like Cairo; it retained a part of the fish that the residents ate; moreover today, the accumulation of silts weakens the dam and causes new technical problems. That does not mean that the Aswan dam should not have been built, but that all the decisions taken in a techno-economic context are likely to be disastrous by their consequences. 

It is like the deviation of rivers in Siberia that the Soviet government made and where the perverse consequences are more important than the positive ones.

It is thus necessary to recognize the inseparability of the separable, at the historical and social levels, as it has been recognized at the microphysical level. According to quantum physics, confirmed by Aspect's experiments, two microphysical entities are immediately connected one to the other although they are separated by space and time. Even more, one arrives to the idea that everything that is separated is at the same time inseparable.

\section{The hologrammatic and dialogical principles}

 The hologrammic or hologrammatic principle should also be advanced, according to which not only a part is inside a whole, but also the whole is inside the part; just as the totality of the genetic inheritance is found in each cell of our organism, the society with its culture is inside the spirit of an individual.
 
 We return again to the \textit{logical core} of complexity which we will see, is dialogical: separability-inseparability, whole-parts, effect-cause, product-producer, life-death, homo sapiens-homo demens, etc.

It is here that the principle of the excluded middle reveals its limit. The excluded middle states ``A cannot be A and not A", whereas it can be one and the other. For example, Spinoza is Jewish and non-Jewish, he is neither Jewish, nor non-Jewish. It is here that the dialogic is not the response to these paradoxes, but the means of facing them, by considering the complementarity of antagonisms and the productive play, sometimes vital, of complementary antagonisms.

\section{For the sciences, a certain number of consequences}

Regarding sciences, we can see a certain number of consequences.

First of all, classical science is somehow complex, even when it produces simplifying knowledge. Why?

Because science is a quadruped which walks on the following four legs: the leg of empiricism made of data, experimentation or observation; the leg of rationality, made of logically constituted theories; the leg of verification, always necessary; and the leg of imagination, because great theories are products of a powerful creative imagination. Thus science is complex, produced by a quadruped movement, which prevents it from solidifying.

The objective knowledge which is its idea, resulted in the need of eliminating subjectivity, i.e. the emotional part inherent to each observer, to each scientist, but it also comprised the elimination of the subject, i.e. the being which conceives and knows. However, any knowledge, including objective, is at the same time a cerebral translation starting from data of the external world and a mental reconstruction, starting from certain organizing potentialities of the spirit. It is certain that the idea of a pure objectivity is utopian. Scientific objectivity is produced by beings who are subjects, within given historical conditions, starting from the rules of the scientific game. The great contribution of Kant was to show that the object of knowledge is co-constructed by our spirit. He indicated us that it is necessary to know knowledge to know its possibilities and limits. The knowledge of knowledge is a requirement of the complex thinking.

As Husserl indicated in the 30's, in particular in his conference on the crisis of European science, sciences developed extremely sophisticated means to know external objects, but no means to know themselves. There is no science of science, and even the science of science would be insufficient if it does not include epistemological problems. Science is a tumultuous building site, science is a process that could not be programmed in advance, because one can never program what one will find, since the characteristic of a discovery is its unexpectedness. This uncontrolled process has lead today to the development of potentialities of destruction and of manipulation, which must bring the introduction into science of a double conscience: a conscience of itself, and an ethical conscience.

Also, I believe that it will be necessary to arrive more and more to a scientific knowledge integrating the knowledge of the human spirit to the knowledge of the object which this spirit seizes and recognizing the inseparability between object and subject.

\section{Two scientific revolutions introduced complexity \textit{de facto}}

I already indicated how the concept of complexity emerged in a marginal fashion in a sphere of mathematicians/engineers. It should be indicated now that the XX$^{th}$ century knew two scientific revolutions which \textit{de facto} introduced complexity without, however, recognizing this notion that remains implicit.

The first revolution, after the thermodynamics of the XIX$^{th}$ century, is that of the microphysics and cosmophysics that introduced indeterminism, risk---where determinism reigned---and elaborated suitable methods to deal with the uncertainties met.

The second revolution is that which gathers disciplines and restores between them a common fabric. It begins in the second half of the XX$^{th}$ century. Thus in the 60's, Earth sciences designed Earth as a complex physical system, which makes it possible today to articulate geology, seismology, vulcanology, meteorology, ecology, etc. At the same time, ecology develops as a scientific knowledge bringing together data and information coming from different physical and biological disciplines in the conception of ecosystems. It makes it possible to conceive how an ecosystem either degrades, develops, or maintains its homeostasis. From the 70's, the ecological conception extends to the whole biosphere, necessarily introducing knowledge from the social sciences.

Although ecology, at the biosphere level, cannot make rigorous predictions, it can give us vital hypothesis, concerning, for example, global warming, which manifests itself by the melting of glaciers in the Antarctic or the Arctic. Thus ecology, cosmology, and Earth sciences have become poly-disciplinary sciences, even transdisciplinary.  Sooner or later, this will arrive in biology, from the moment when the idea of self-organization will be established; this will arrive in the social sciences\footnote{Cf my \textit{Humanit\'{e} de l'humanit\'{e}, La  M\'ethode 5}, Le Seuil.}, although they are extremely resistant.

Finally, the observer, chased by the objectivity postulate, was introduced into certain sciences, such as microphysics where the observer perturbs what it observes. In the case of cosmology, even if one does not adhere to what Brandon Carter called the anthropic principle, which holds account of the place of humans in the universe, one is obliged to conceive that this universe, among its perhaps negligible possibilities, had the possibility of human life, perhaps only on this planet Earth, but perhaps also elsewhere.

Thus, the common fabric between the human, the living, and the Universe can be restored, which implies a complex conception capable at the same time to distinguish the human from the natural and to integrate it.

\section{The insertion of science in History}

In addition, there is the problem of the insertion of the Sciences in human History.

You know that there are two conceptions of history of sciences, the internalist conception and the externalist conception. The internalist mode sees the development of sciences in isolation, only in function of their internal logic and their own discoveries. The externalist mode sees them in function of historical and social developments which determine the scientific developments.

I think that it is necessary to link both, and this is appropriate for other developments than those of sciences. Thus, some wanted to understand the perversion of the Soviet Union starting from internal factors, such as insufficiencies of the Marxist doctrine, limitations of that of Lenin. Others wanted to impute it to external elements such as the surrounding and hostility of the capitalist powers with regard to the Soviet Union or former elements such as the backwardness of tsarist Russia. Whereas the true cognitive game is to link these two aspects in a dialogical fashion.

If one continues to place oneself from the viewpoint of modern Western history of science, one sees how from its marginal and quasi-deviating birth in the XVII$^{th}$ century, it is developed in the XVIII$^{th}$, introduced in universities in the XIX$^{th}$, then in states and companies in the XX$^{th}$, and how it becomes central and driving within human history in the form of techno-science, and produces not only all the major elements for a renewed knowledge of the world and beneficial effects for humanity, but also formidable and uncontrolled powers which threaten it.

I don't know if I am right or wrong in retaking an expression of Vico, but it is necessary for us to arrive to the ``Scienza Nuova". Very precisely, Vico inscribed the historical perspective at the heart of the scienza nuova. It is necessary to amplify the idea of scienza nuova by introducing the interaction between the simple and the complex, by conceiving a science that does not suppress disciplines but connects them, and consequently makes them fertile, a science which can at the same time distinguish and connect and where the transdisciplinarity is inseparable from complexity.

I repeat it, as much as the compartmentalization of disciplines disintegrates the natural fabric of complexity, as much a transdisciplinary vision is capable of restoring it.

\section{The link between science and philosophy}

The link between science and philosophy has been broken. Still in the XVII$^{th}$ century, the great scientists were at the same time great philosophers. Certainly, they did not identify Science and Philosophy. When Pascal made his experiments in Puy de D\^ome, he did not think about the bet problem. But in the times of Pascal, Gassendi, Leibniz, there was not this cut. This became a frightening ditch. The ditch of ignorance separates the scientific culture from the culture of the humanities.

But the current has started to be reversed: the most advanced sciences arrive to fundamental philosophical problems: Why is there a universe out of nothing? How was this universe born from a vacuum which was not at the same time the vacuum? What is reality? Is the essence of the universe veiled or totally cognizable?

The problem of life is posed from now on in a complexity that exceeds biology: the singular conditions of its origin, the conditions of emergences of its creative powers. Bergson was mistaken by thinking that there was an \textit{\'elan vital}, but was right while speaking about creative evolution. He could even have spoken about evolutionary creativity.

Today we can foresee the possibility of creating life. From the moment when it is believed that life is a process developed starting only from physicochemical matter under certain conditions, in underwater thermal vents or elsewhere, one can very well consider creating the physical, chemical, thermodynamic conditions which give birth to organisms gifted with qualities that one calls life. We can also foresee the possibility to modify the human being in its biological nature. Therefore, we have to meditate about life, as we never did it. And at the same time we must meditate about our relationship with the biosphere.

Thus all the most advanced sciences arrive to fundamental philosophical problems that they thought to have eliminated. They do not only find them, they renew them.

If one defines philosophy by the will and capacity of reflection, it is necessary that the reflectivity is also introduced into the sciences, which does not eliminate the relative autonomy of philosophy nor the relative autonomy of scientific procedures compared to philosophical procedures.

Finally and especially, any knowledge, including the scientific one, must comprise in itself an epistemological reflection on its foundations, principles, and limits.

Still today there is the illusion that complexity is a philosophical problem and not a scientific one. In a certain way, it is true, in a certain way, it is false. It is true when you place yourselves from the point of view of an isolated and separated object: the fact that you isolate and separate the object made the complexity to disappear: thus it is not a scientific problem from the point of view of a closed discipline and a decontextualized object. But, as soon as you start to connect these isolated objects, you are in front of the problem of complexity.

\section{Second epistemological rupture with restricted complexity}

It is here that a second epistemological rupture with restricted complexity appears.

Restricted complexity is interested  essentially in dynamical systems called complex. That is to say, it constitutes its own field, within the field of sciences.

But generalized complexity not only concerns all fields, but also relates to our knowledge as human beings, individuals, persons, and citizens. Since we have been domesticated by our education which taught us much more to separate than to connect, our aptitude for connecting is underdeveloped and our aptitude for separating is overdeveloped; I repeat that knowing, is at the same time separating and connecting, it is to make analysis and synthesis. Both are inseparable, and our atrophy of the capacity to connect is increasingly serious in a globalized, complexified mode, where it is a matter of generalized interdependence of everything and everyone.

The International Ethical, Political and Scientific Collegium has formulated a declaration of interdependence which it would wish to see promulgated by the United Nations. We must think the interdependence in all fields, including the complex relation between the parts and the whole. We need to be able to face uncertainties of life whereas nothing prepares us for it. We need to face complexity, including for action, whereas one opposes the cautionary principle to the risk principle, while Pericles had truly expressed the union of the two antagonistic principles when he said during a speech to the Athenians during the Peloponnesian war: ``we Athenians, we are capable of combining prudence and audacity, whereas the others are either timorous or bold". It is the combination which we need. Also, precaution needs today sometimes much invention.

We need to deeply reform all our way of knowing and thinking.

\section{The principle of ecology of action}

The principle of ecology of action is, in my opinion, central: from the moment an action enters a given environment, it escapes from the will and intention of that which created it, it enters a set of interactions and multiple feedbacks and then it will find itself derived from its finalities, and sometimes to even go in the opposite sense. The ecology of action has a universal value, including for the development of sciences, whose destructive nuclear consequences were absolutely unexpected.

Think that when Fermi elucidated the structure of the atom in the 30's, it was a purely speculative discovery and he had by no means thought that this could allow the fabrication of an atomic bomb. However, a few years later, the same Fermi went to the United States to contribute to the fabrication of the atomic bomb that would be used in Hiroshima and Nagasaki. When Watson and Crick determined the structure of the genetic inheritance in DNA, they thought that it was a great conquest of knowledge without any practical consequences. And hardly ten years after their discovery, the problem of genetic manipulations was posed in the biology community.

The ecology of action has a universal value. One can think of examples in our recent French history: a dissolution of the Parliament by President Chirac to have a governmental majority led to a socialist majority; a referendum made to win general support led to its rejection. Gorbachev tried a reform to save the Soviet Union but this one contributed to its disintegration. When one sees that a revolution was made in 1917 to suppress the exploitation of man by his fellow man, to create a new society, founded on the principles of community and liberty, and that this revolution, not only caused immense losses of blood, destruction, and repression by a police system, but, after seventy years, it led to its contrary, i.e. to a capitalism even more fierce and savage than that of the tsarist times, and with a return of religion! Everything that this revolution wanted to destroy resurrected. How not to think about the ecology of action!

\section{Creating ``Institutes of fundamental culture"}

The reform of the spirit seems to me absolutely necessary.

Once that I had understood that the reform of thought, deep work that I carried out in \textit{La M\'ethode}, is a necessity, I accepted the offer of a Minister of Education when he called me for the reform of the content of secondary education. I tried to introduce my ideas of reform of thought into an educational project. I saw its total failure---finally it did not failed, it was not applied!---That pushed me to reflect even more. I wrote a book called 
\textit{La T\^ete bien faite} (The head well made), then on the initiative of UNESCO I made a book called \textit{Les Sept savoirs n\'{e}cessaires \`a l'\'{e}ducation du futur} (The seven knowledges necessary in the education of the future).

Following a University which will be created on these principles in Mexico, I had the more restricted but maybe more necessary idea of creating ``Institutes of fundamental culture", which would be sheltered in a University or independent, addressing everybody, i.e. before University or during University or after University, students, citizens, members of trade unions, entrepreneurs, everybody.

Why the word ``fundamental culture"? Because it is that what is missing. In fact, it is the most vital matter to be taught, the most important to face life, and which is ignored by education.

\begin{enumerate}
\item Knowledge as a source of error or illusion; nowhere the traps of knowledge are taught, which come owing to the fact that all knowledge is translation and reconstruction.

\item Rationality, as if it were an obvious thing, whereas we know that rationality knows its perversion, its infantile or senile diseases.

\item Scientificity. What is science, its frontiers, its limits, its possibilities, its rules. Moreover, there is an abundant literature, but which has never been consulted by the scientists who are recruited at CNRS for example. Most of the time, they do not know anything about the polemic between Niels Bohr and Einstein, the works of Popper, Lakatos, Kuhn, etc.

\item What is complexity.

\end{enumerate}

And also:

\begin{itemize}
\item A teaching on ``what is the human identity and condition", which is not found anywhere.

\item A teaching on the global age, not only today's globalization, but all its antecedents starting from the conquest of America, the colonization of the world, its current phase, and its future prospects.

\item A teaching on human understanding.

\item A teaching concerning the confrontation of uncertainties in all the fields: sciences, everyday life, history (we have lost the certainty of progress, and the future is completely uncertain and obscure).

\item A teaching on the problems of our civilization.

\end{itemize}

That is for me the fundamental teaching that can aid the reform of the spirit, of thought, of knowledge, of action, of life.

\section{I conclude: generalized complexity integrates restricted complexity}

Unfortunately, restricted complexity rejects generalized complexity, which seems to the former as pure chattering, pure philosophy. It rejects it because restricted complexity did not make the epistemological and paradigmatic revolution which complexity obliges. That will be done without a doubt. But in the meantime, we see that the problematic of complexity have invaded all our horizons, and I repeat ``problematic", because it is an error to think that one will find in complexity a method that can be applied automatically to the world and anything.

Complexity is a certain number of principles which help the autonomous spirit to know. Whereas a program destroys the autonomy of the one who seeks, the problematic of complexity stimulates an autonomous strategy, obliges in the field of action---once that one knows that the ecology of action can pervert the best intentions---to reconsider our decisions like bets and incites us to develop an adequate strategy to more or less control the action.

In other words, in all the fields, I would say ``help yourself and the complexity will help you", which has nothing to do with the mechanical application of a program or a rule. It is a deep reform of our mental functioning, of our being.

These ideas now marginal, deviating, begin to constitute a tendency still in minority, or rather tendencies since there are several paths to go towards complexity. These ideas, these deviations, can be developed and become cultural, political, and social forces.

The probabilities of a global future are extremely alarming: our spaceship is pulled by four engines without any control: science, technology, economy, and the search for profit---all this under conditions of chaos since the techno-civilizational unification of the planet, under the Western push, causes singular cultural resistances and cultural and religious re-closings.

The planet is in crisis with all the possibilities, ones regressive and destructive, others stimulant and fertile, such as invention, creation, new solutions.

\section{We should even apprehend the possibilities of metamorphosis}

We should even apprehend the possibilities of metamorphosis because we have completely astonishing examples of it from the past. The change in certain places where there have been demographic concentrations in the Middle East, in the Indus basin, in China, in Mexico, in Peru, from prehistoric societies of hundreds of men, without cities, without state, without agriculture, without army, without social class, to enormous historical societies with cities, agriculture, army, civilization, religion, philosophy, works of art... that constituted a sociological metamorphosis.

Perhaps we are going towards a meta-historical metamorphosis suitable for the birth of a society-world at a global scale.

I would say that complexity does not put us only in the distress of the uncertain, it allows us to see besides the probable, the possibilities of the improbable, because of those which have been in the past and those that can be found again in the future.

We are in an epoch of doubtful and uncertain combat.

That makes one think of the Pacific war, after the Japanese had broken into the Pacific Islands and had begun to threaten California, there was a gigantic naval fight over 200 kilometers along the Midways between the Japanese and American fleets: battleships, aircraft carriers, submarines, planes. The global vision was impossible for both of them: there were sunken Japanese ships, sunken American ships, planes that did not find the enemy fleet; in short, total confusion, the battle divided in several fragments. At a given moment, the Japanese Admiral realizing his losses in battleships and planes, thought that they were defeated, thus called for retreat. But the Americans, who had lost as much, were not the first to think that they were defeated; after the Japanese retreat, they were victorious.

Well, the outcome of what will happen, we cannot conceive it yet! We can always hope and act in the direction of this hope.

The intelligence of complexity, isn't it to explore the field of possibilities, without restricting it with what is formally probable? Doesn't it invite us to reform, even to revolutionize?

\end{document}